\documentclass[11pt,a4paper]{article}
\usepackage[hyperref]{emnlp2020}
\usepackage{times}
\usepackage{latexsym}
\usepackage{graphicx}

\newcommand{\eat}[1]{}

\usepackage{microtype}

\aclfinalcopy 

\title{Using the Hammer Only on Nails:\\A Hybrid Method for Evidence Retrieval for Question Answering}

\author{Zhengzhong Liang$^1$, Yiyun Zhao$^2$, Mihai Surdeanu$^1$ \\
  $^1$ Computer Science Department, The University of Arizona \\
  $^2$Linguistics Department, The University of Arizona \\
  Tucson, AZ 85718 \\
  \texttt{$\{$zhengzhongliang, yiyunzhao, msurdeanu$\}$@email.arizona.edu} \\
  }

\date{}

\begin{document}
\maketitle
\begin{abstract}
Evidence retrieval is a key component of explainable question answering (QA). 
We argue that, despite recent progress, transformer network-based approaches such as universal sentence encoder (USE-QA) do not always outperform traditional information retrieval (IR) methods such as BM25 for evidence retrieval for QA. 
We introduce a lexical probing task that validates this observation: we demonstrate that neural IR methods have the capacity to capture lexical differences between questions and answers, but miss obvious lexical overlap signal. 
Learning from this probing analysis, we introduce a hybrid approach for evidence retrieval that combines the advantages of both IR directions.
Our approach uses a routing classifier that learns when to direct incoming questions to BM25 vs. USE-QA for evidence retrieval using very simple statistics, which can be efficiently extracted from the top candidate evidence sentences produced by a BM25 model. 
We demonstrate that this hybrid evidence retrieval generally performs better than either individual retrieval strategy on three QA datasets: OpenBookQA, ReQA SQuAD, and ReQA NQ. Furthermore, we show that the proposed routing strategy 
is considerably faster than neural methods, with a runtime that is up to 5 times faster than USE-QA.
\end{abstract}

\section{Introduction}
Open-domain question answering  (QA) systems traditionally have three components: evidence retrieval, evidence reranking, and answer classification/extraction. In evidence retrieval, the model retrieves a smaller set of possibly useful evidence texts from a large knowledge base (KB), which are then reranked by the following component to push the most useful information to the top. 
Traditional directions use word-overlap based models for evidence retrieval such as tf-idf and BM25. However, this can potentially cause the missing of useful information due to the ``lexical chasm''~\cite{berger2000bridging} between the question and the answer. A potential remedy for this is to use neural networks for evidence retrieval, such as transformer network-based contextualized embedding methods~\cite{devlin2019bert, yang2019xlnet, yang2019multilingual}. 


Focusing on this evidence retrieval stage of a QA system,
we argue that, for this component, transformer networks should not always be preferred over standard information retrieval (IR) methods. While transformer networks have the advantage that they can capture and use semantic information, 
they have drawbacks as well. 
First, due to their reliance on continuous representations, these methods do not take direct advantage of obvious lexical evidence. This is a drawback in long-text retrieval, which tends to be affected less by the lexical chasm problem than short-text retrieval. 
Second, transformer-based methods are expensive to run, which makes them a less than ideal choice for end-user NLP applications with temporal constraints. 

In this paper we introduce a {\em hybrid} approach for evidence retrieval for question answering. 
Our approach uses a routing classifier that routes an incoming question to either an IR method or a supervised transformer method for evidence retrieval, using solely shallow statistics sampled from the knowledge base of explanatory texts for each question. This strategy has two benefits: first,  evidence retrieval performance improves overall because each question is handled by the appropriate retrieval method. Second, this method reduces computational overhead because for a considerable number of questions it does not use the more expensive neural component. 

In particular, our contributions are:

{\flushleft {\bf (1)}} 
We design and conduct a series of supervised lexical probing tasks on two QA datasets, which are trained to predict the terms in the query and the gold evidence text from the entire vocabulary, using as input either the tf-idf vector of the query, or the neural embedding of the same query. 
The comparison of the two probes indicates that the probe trained from the tf-idf vector of the query tends to predict terms that exist in the original query (thus emphasizing lexical overlap), whereas the probe trained on top of the query's neural embedding predicts more terms in the evidence text that do not exist in the query (thus bridging the lexical chasm).
This validates our hypothesis that different retrieval strategies should be used in different scenarios.

{\flushleft {\bf (2)}} 
Learning from this observation, we propose a hybrid retrieval method, which routes queries to either an information retrieval method (BM25 \cite{bm25ref}) or a transformer-based one (USE-QA \cite{yang2019multilingual}).
We show that this routing decision is learnable from simple statistics that can be efficiently extracted from the top documents retrieved by an IR method.

{\flushleft {\bf (3)}} 
We show that using this hybrid strategy generally improves evidence retrieval performance in three QA datasets: OpenBookQA \cite{mihaylov2018can}, ReQA SQUAD, and ReQA Natural Questions (NQ) \cite{ahmad2019reqa}. The hybrid approach performs significantly better than either individual model on ReQA SQUAD and NQ, with improvements in the mean reciprocal rank (MRR) of the correct evidence sentence ranging from  1\% to 7.4\% (depending on the dataset). On OpenBookQA the difference between the hybrid method and USE-QA is not statistically significant. 

{\flushleft {\bf (4)}} 
Our analysis indicates that the hybrid method is significantly faster than neural IR methods. For example, the hybrid method is 2.2 times faster than USE-QA in OpenBookQA, and 5.2 times faster in ReQA SQUAD. 

\section{Related Work}
Neural IR methods 
provide an exciting potential direction to mitigate the lexical chasm in QA \cite{dehghani2017neural}. Neural IR approaches can be broadly divided into two categories: representation-based and interaction-based \cite{guo2019deep}. Representation-based neural IR directions {\em pre-encode} the query and the document into a continuous representation learned using a subsample of the data, and use a shallow method to compute relevance scores at runtime (e.g., dot product)~\cite{huang2013learning, lee2019latent}. Representation-based neural IR methods have low runtime overhead because all documents can be pre-computed as vectors, so that at test time the embeddings of the documents do not have to be recomputed for each query (i.e., the neural model is run for $N_q$ times at test time, where $N_q$ is the number of queries). 

Interaction-based methods learn a query-specific representation of the documents {\em at runtime} \cite{hu2014convolutional,qiao2019understanding,nogueira2019passage}. Usually the query and the candidate document are concatenated and processed by a neural model jointly, so that complex interactions of the terms in the query and the document can be better captured. However, this requires running the neural model for $N_q \cdot N_d$ times at test time (where $N_q$ is the number queries and $N_d$ is the number of docs). Therefore interaction-based methods are not suitable for large-scale first stage retrieval and are usually used for second-stage retrieval (reranking). In this paper we focus on the representation-based method in the first stage retrieval.

Empirical evidence has shown that neural IR methods perform better in short-text retrieval, where the word-overlap-based IR methods are more likely to suffer from the lexical chasm problem. However, not much work has been done to show why neural IR methods are able to reduce the lexical chasm problem \cite{guo2019deep}, partly because it is hard to explain the meaning of neural embeddings. Recently, probing tasks have been widely used to help understand the properties of neural networks \cite{conneau2018you, hewitt2019structural, hewitt2019designing}. In probing tasks, a shallow model is placed on top of the large neural model, and the shallow model is trained to show some properties of the large model. For example, in \citet{hewitt2019structural}, the authors show that some syntactic information is encoded in the embeddings of the intermediate layers of BERT. Inspired by this, we design and conduct a series of lexical probing tasks to compare the abilities of traditional IR methods and neural IR methods to predict the terms that are indicative of lexical chasm, i.e., they exist in the evidence sentences but not in the original query.

\begin{table*}[t!]
\centering
\small
\begin{tabular}{p{2.3cm} p{1.5cm} p{1.5cm} p{1.3cm} p{1.3cm} p{2.5cm}  p{2.5cm}}
\hline \textbf{Dataset} & \textbf{N train} & \textbf{N dev}  & \textbf{N test}  & \textbf{N doc} & \textbf{Avg. query len.} & \textbf{Avg. doc. len.} \\ \hline
OpenBookQA & 4,957  & 500 & 500 & 1,326 & 13.71 & 9.49 \\
ReQA SQuAD  & 87,599  & 11,426 & N/A & 101,957 & 10.38 & 160.62\\
ReQA NQ      & N/A  & N/A & 74,097 & 239,013 & 9.09 & 146.16 \\
\hline
\end{tabular}
\caption{Statistics of the three datasets used throughout this paper, including the number of queries in the train/dev/test set, the number of candidate documents, and the average number of tokens of each query/document.}
\label{tab:dataset_intro}
\end{table*}

\begin{table*}[t!]
\centering
\small
\begin{tabular}{p{2.3cm} p{3cm} p{4.0cm} p{4.8cm} } \hline
\textbf{Dataset} & \textbf{Query} & \textbf{Answer Sentence}  & \textbf{Context}   \\ \hline
ReQA SQuAD   &  To whom did the Virgin Mary allegedly appear in 1858 in Lourdes France?  &   It is a replica of the grotto at Lourdes, France where the Virgin Mary reputedly appeared to Saint Bernadette Soubirous in 1858.  & ... a Marian place of prayer and reflection. It is a replica of the grotto at Lourdes, France where the Virgin Mary reputedly appeared to Saint Bernadette Soubirous in 1858. At the end of the main drive ...  \\
ReQA NQ  &  Who sings the song i don't care i love it   &  In its chorus, Icona Pop and Charli XCX shout in unison ``I don't care / I love it''. &  ...  breaking up with an older boyfriend. In its chorus, Icona Pop and Charli XCX shout in unison ``I don't care / I love it''. Critics compared the song's breakup ... \\
OpenBookQA & Tadpoles start their lives as Water animals  &  Tadpole changes into a frog  & N/A\\
\hline
\end{tabular}
\caption{Examples of queries, answer sentences, and contexts in the three datasets.}
\label{tab:dataset_intro_example}
\end{table*}

%
Although they do not rely on explicit word overlaps, neural IR methods do not always outperform traditional IR. For instance, it has been shown that neural IR models usually work better on short text retrieval \cite{cohen2016adaptability}, and when training data is abundant \cite{guo2019deep}, but not in other situations \cite{iyyer2015deep}.

Efforts have been made to combine neural IR with traditional IR in open domain QA. For example, multiple approaches use traditional IR for evidence retrieval and neural IR for evidence reranking \cite{yang2019end, pirtoacua2019answering, nie2019revealing, chen2015harnessing}. However, {\em always} relying on traditional IR for evidence retrieval may miss useful evidence that does not have large lexical overlap with the query.

Motivated by these works, in this paper we propose a  hybrid evidence retrieval direction for first-stage retrieval, in which we learn {\em when} to use traditional IR vs. neural IR. As our results show, this yields a more accurate retrieval component that also has a lower runtime overhead than neural methods.

\section{Datasets and Evaluation Measures}
\label{sec:approach_dataset}

We conduct our probing analyses and retrieval experiments on three QA-related retrieval datasets. 
One of these datasets comes from the science domain; the other two are open domain.
More statistics and examples of these datasets are shown in table \ref{tab:dataset_intro} and \ref{tab:dataset_intro_example}, and we describe them below.

{\flushleft {\bf OpenBookQA:}} The OpenBookQA dataset \cite{mihaylov2018can} addresses a multiple-choice QA task in the science domain. Each correct answer is jointly annotated with one key evidence sentence (or {\em justification}) that supports its correctness. The justification comes from a knowledge base of 1326 sentences. In this paper, we construct a corpus of 1326 documents from these sentences. Further, for each question,
we concatenate the question and the correct answer choice to form the {\em query}, and retrieve the gold justification (or {\em target document}) for that query from the corpus of 1326 documents.

{\flushleft {\bf ReQA SQuAD:}} The ReQA SQuAD dataset \cite{ahmad2019reqa} is a sentence-level retrieval dataset converted from the SQuAD reading comprehension dataset \cite{rajpurkar2016squad}. In the original SQuAD reading comprehension task, 
the answers to questions must be extracted from sentences in a set of provided paragraphs. 
The ReQA SQuAD dataset uses the questions in SQuAD as the queries, and converts all paragraphs to single sentences. The goal of this retrieval task is  to retrieve the sentence that contains the correct answer from all the sentences generated from all the paragraphs. Since some answer sentences are meaningless without the surrounding context, each candidate sentence is accompanied by its original paragraph as the context.

{\flushleft {\bf ReQA NQ:}} The ReQA NQ dataset \cite{ahmad2019reqa} is similarly converted from another reading comprehension task -- Natural Questions \cite{kwiatkowski2019natural} --  following the same process as ReQA SQuAD. Similarly, each query is a question and each target document is a sentence/context pair, where the context is the paragraph that contains the gold justification. 

\section{Understanding the Behavior of Neural IR through Lexical Probing}
\label{sec:probe}

\begin{table*}[ht]
\centering
\small
\begin{tabular}{lccccc}
\hline \textbf{Dataset}& \textbf{Task} & \textbf{Query MAP}  & \textbf{Query PPL} & \textbf{Fact MAP}  & \textbf{Fact PPL} \\ \hline
OpenBookQA & USE-QA embd, gold label  & 0.306 {\tiny $\pm0.007$} & 1.709 {\tiny $\pm0.020$} & 0.154 {\tiny $\pm0.006$} & 1.188 {\tiny $\pm0.011$} \\
                       & tf-idf embd, gold label & 0.458 {\tiny $\pm0.008$} & 1.558 {\tiny $\pm0.006$} & 0.098 {\tiny $\pm0.004$} & 1.334 {\tiny $\pm0.014$} \\
                       & Random embd, gold label &  0.053 {\tiny $\pm0.015$} & 3.640 {\tiny $\pm0.097$} & 0.031 {\tiny $\pm0.007$} & 3.294 {\tiny $\pm1.626$} \\
                       &  USE-QA embd, rand label & 0.085 {\tiny $\pm0.003$} & 1.974 {\tiny $\pm0.016$} & 0.043 {\tiny $\pm0.002$} & 1.557 {\tiny $\pm0.015$} \\ \hline
ReQA SQuAD  &  USE-QA embd, gold label  & 0.139 {\tiny $\pm0.005$} & 1.944 {\tiny $\pm0.009$} & 0.147 {\tiny $\pm0.001$} & 2.134 {\tiny $\pm0.001$} \\
                     &  tf-idf embd, gold label  & 0.142 {\tiny $\pm0.041$} & 1.828 {\tiny $\pm0.003$} & 0.127 {\tiny $\pm0.007$} & 2.043 {\tiny $\pm0.001$} \\
                     &  Random embd, gold label  & 0.091 {\tiny $\pm0.004$} & 7.419 {\tiny $\pm3.498$} & 0.093 {\tiny $\pm0.002$} & 3.478 {\tiny $\pm0.145$} \\
                    &  USE-QA embd, rand label & 0.122 {\tiny $\pm0.020$} & 1.909 {\tiny $\pm0.004$} & 0.124 {\tiny $\pm0.004$} & 2.013 {\tiny $\pm0.001$} \\
\hline
\end{tabular}
\caption{Results of the probing tasks on two datasets: OpenBookQA (science domain) and ReQA SQuAD (open domain). We report mean average precision (MAP) (higher is better) and perplexity (PPL) (lower is better) scores for the gold terms to be predicted. We report separate scores for terms in the query, and terms that occur {\em only} in the justification fact. We report the mean and standard deviation across 5 random seeds. 
}
\label{tab:embd_probe}
\end{table*}

Our key hypothesis is that neural IR methods are better at modeling the lexical chasm between queries and evidence sentences than traditional IR, whereas traditional IR captures explicit lexical overlap better. We design a lexical probe and control tasks to investigate this.
\subsection{Task Overview}
\begin{figure}
\centering
\includegraphics[width = 0.48\textwidth]{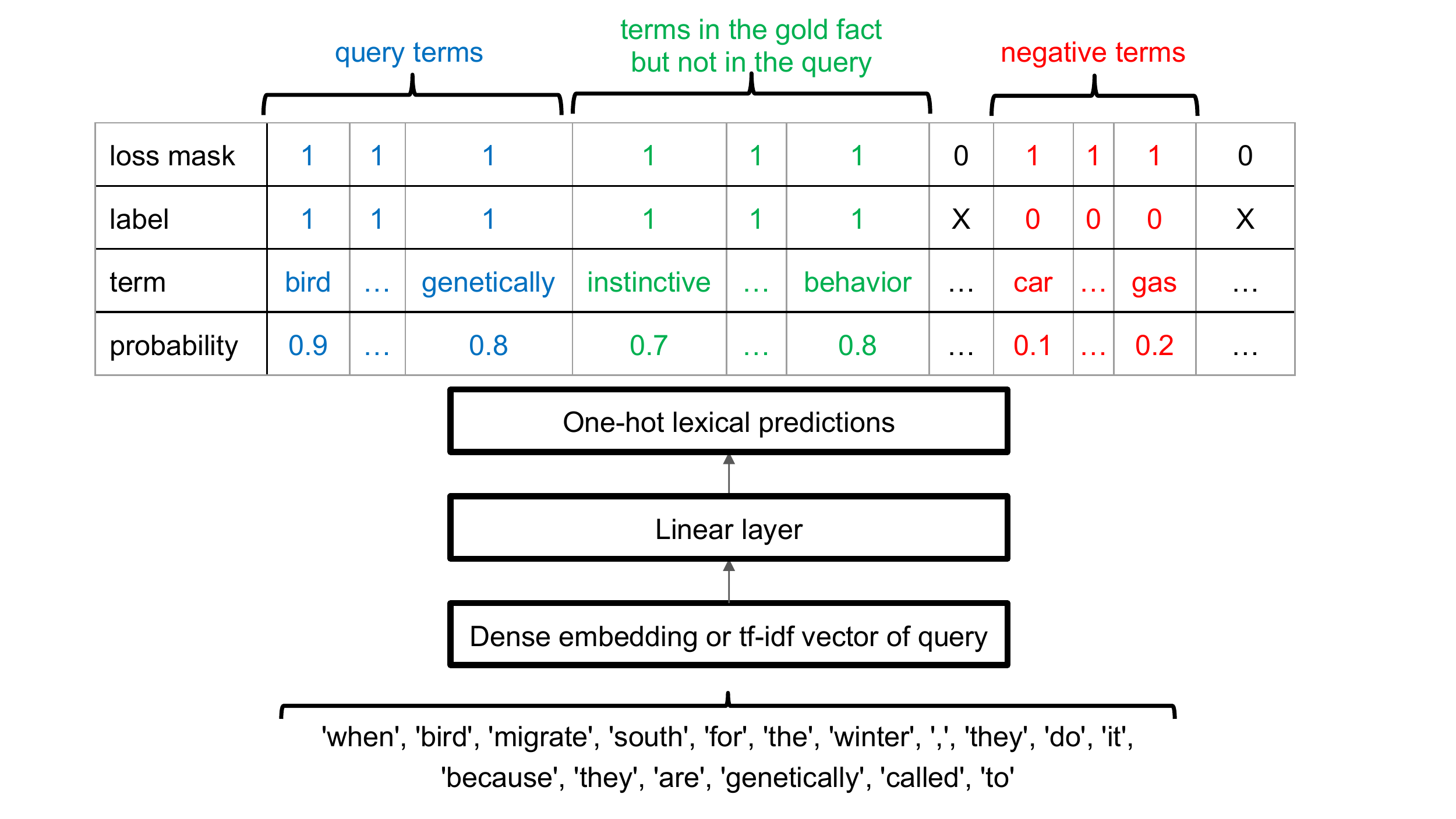}
\caption{\footnotesize Probe task overview: the linear probe is trained to predict the terms in the query and \eat{or \em{only}}in the gold fact from the entire vocabulary, given either the input embedding or the tf-idf vector of the query. This probe investigates the capability of the query representation to predict both lexical overlap (i.e., terms from the query), as well as its ability to bridge the lexical chasm between queries and supporting facts (i.e., predict terms that exist in the fact and not in the query). A loss mask is used to make sure the loss is only computed on certain terms during training.
}
\label{fig:fig_1_probe}
\end{figure} 
Figure \ref{fig:fig_1_probe} summarizes our lexical probe, with a walkthrough example from OpenbookQA. 
{\flushleft {\bf Probe input:}} The probe starts by generating a representation of the input query. This representation is either: (a) the tf-idf vector of the query, generated using \textit{scikit-learn} \cite{scikit-learn}, 
or (b) the query embedding generated by USE-QA \cite{yang2019multilingual}. 
{\flushleft {\bf Linear layer:}} This vector is fed to a linear layer, with input size $N_d$ and output size $N_v$, to predict the terms (i.e., unique words) in the query {\em and} in the gold fact. $N_v$ is the size of vocabulary $\mathcal{V}$, where $\mathcal{V}$ is the set of all terms in the dataset.  Each number in the output is the predicted probability of that particular term being in the query/gold fact.\footnote{A sigmoid function is applied to the raw output of the linear score to normalize each score to the scale 0 to 1. More details can be found in Appendix \ref{sec:supp_probe}.} Note that the input embedding/vector is not changed during the training of the probe task. Thus, if the neural embedding contains meaningful information about the gold fact, it should perform better than tf-idf on predicting the terms that are only in the gold fact.
\vspace{-1mm}{\flushleft {\bf Training label and loss:}} For each query $q_i$, we use $\mathcal{P}_i$ to indicate the set of all terms in $\{q_i, \textrm{gold\_fact}(q_i)\}$. The training label is a one-hot vector of size $N_v$, where the values for the terms in $\mathcal{P}_i$ are 1, and the rest of the entries are 0. However, since the terms in $\mathcal{P}_i$ are considerably fewer than the whole vocabulary, there will be many more 0s than 1s in this label vector, causing label imbalance. Therefore, we construct a set of negative terms $\mathcal{N}_i$, which contains terms that are randomly sampled from the vocabulary $\mathcal{V}$ but not in $\mathcal{P}_i$. The size of $\mathcal{N}_i$ also equals to the size of $\mathcal{P}_i$. The loss is only computed on the terms in $\mathcal{P}_i \cup \mathcal{N}_i$ instead of the whole vocabulary $\mathcal{V}$. The total loss of each query $q_i$ is summarized by: $L =-\sum_{j \in \mathcal{P}_i \cup \mathcal{N}_i} [y_j \log \hat{y_j}+(1-y_j)\log(1-\hat{y_j}) ]$, where $y_j \in \{0, 1\}$ is the label, and $\hat{y_j}\in(0,1)$ is the predicted probability of the corresponding term. 
\subsection{Control Tasks}
\label{sec:probe_control}
To make sure that the information necessary for prediction is contained in the query representation and not in the linear layer \cite{hewitt2019designing}, we designed two control tasks:
{\flushleft {\bf Random embedding (rand embd):}} This probe replaces the neural embedding with a randomly-generated embedding. If the query representation encodes useful information, this probe should perform much worse than the one using the neural representation.
{\flushleft {\bf Rand label (rand label):}} In this experiment we randomly replace the target terms in both training and testing. For example, we replace the terms to be predicted for query $i$ with terms from a randomly-selected query $j$: $\{q_j, \textrm{gold\_fact}(q_j)\}$.
This is to examine whether it is possible for the linear probe to learn non-sensical associations between random (embedding, target terms) pairs. 

\subsection{Probe Results}
Table~\ref{tab:embd_probe} lists the results of these probing tasks. We draw several observations from these experiments:
{\flushleft {\bf (1)}} With minor exceptions, the two actual probes perform better than the two control tasks. This confirms that there is indeed signal that is encoded in the query representations, and this is what the linear probe classifier exploits.

{\flushleft {\bf (2)}} The probe that relies on the neural query representation obtains higher fact MAP (and lower fact PPL) than the probe that uses the tf-idf representation. This indicates that the neural representation does indeed contain information that helps bridge queries, answers, and supporting facts. On the other hand, the tf-idf probe has higher query MAP (and lower query PPL) than the neural probe. This confirms that the traditional IR representation is better at capturing explicit lexical overlap with the query than the neural one. 
All in all, these observations suggest that these two retrieval directions are better at different things. 


\section{Hybrid Retrieval Approach}

\subsection{Individual IR Models}
\label{sec:approach_model}

The hybrid approach proposed builds from (and is compared against) the following individual retrieval models. Note that these approaches were chosen because they had the best performance on these datasets. For example, USE-QA consistently performed better than BERT. 

{\flushleft {\bf BM25:}} We use the Lucene 6.4.0\footnote{\url{https://lucene.apache.org}} Java implementation of BM25 \cite{bm25ref} as the ``traditional'' IR method. For OpenBookQA, each document is one sentence in the knowledge base corpus (1326 sentences in total). In ReQA SQuAD and ReQA NQ, each document is constructed by concatenating the candidate answer sentence and its context (so that each candidate answer sentence appears twice in the document). In OpenBookQA, we found that proper lemmatization improves the retrieval performance; we used the Stanford CoreNLP toolkit \cite{manning2014stanford} to lemmatize all queries and sentences of OpenBookQA.  

{\flushleft {\bf BERT:}} For this method we fine-tune a pretrained BERT-base model~\cite{devlin2019bert, Wolf2019HuggingFacesTS}. We use the BERT retriever in the representation-based manner: the query $q_i$ and the document $d_j$ are encoded using the [CLS] embedding of BERT as $h_i^q$ and $h_j^d$. Then the relevance score of $q_i$ and $d_j$ are obtained by $Rel(q_i,f_j)=h_i^q \cdot h_j^d$. For ReQA SQuAD and ReQA NQ, each document is composed of the candidate answer sentence and its context. We concatenate them and separate them with the [SEP] token. Therefore, the input of each document is ``[CLS] candidate answer sentence [SEP] context sentences [SEP]''. The details of fine-tuning BERT are provided in Supplemental Material \ref{sec:supp_finetune_bert}.

{\flushleft {\bf USE-QA:}} The USE-QA retriever \cite{yang2019multilingual} has separate encoders for the query and document. The query encoder is a transformer-based model, producing a 512-dimension embedding as the query representation. The document encoder has a transformer-based model to encode the answer sentence and a Convolutional Neural Network (CNN)-based model to encode the context. A single 512-dimension embedding is produced as the document representation. Finally, the relevance score is computed as the dot product of the query embedding and the document embedding. USE-QA is pre-trained on large scale retrieval tasks and, as used in \cite{ahmad2019reqa}, we do not fine-tune it in the retrieval tasks.

\subsection{Are Neural IR Methods Generally Better than Traditional IR?}

The probe introduced in Section~\ref{sec:probe} indicates that neural and traditional IR methods have different behaviors. But what impact does that have in practice, with respect to overall performance? To answer this question, we performed a comparison that aims to understand if transformer-based retrieval methods are better overall than traditional IR. Due to space limitations, we discuss here results from the best individual models in each class: BM25 for traditional IR, and USE-QA for neural IR.\footnote{We observed similar behavior from tf-idf and BERT.} We use two datasets: one domain-specific (OpenBookQA) and one open-domain (SQuAD). 

\begin{figure}
\centering
\includegraphics[width = 0.45\textwidth]{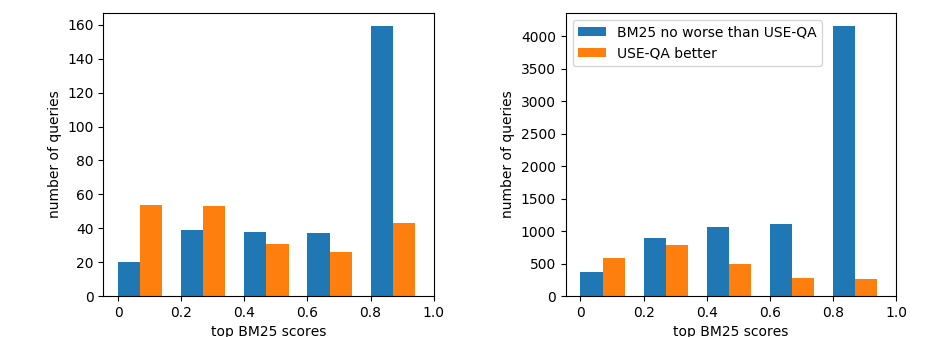}
\caption{\footnotesize Histograms of queries in OpenBookQA dev (left) and a randomly sampled subset of ReQA SQuAD train (right) where BM25 is no worse than USE-QA (blue) or where USE-QA is better (orange). The $x$ axis is the top BM25 score after a softmax is applied to the BM25 scores of the top 64 sentences.
}
\label{fig:fig_0_histogram}
\end{figure}

Figure \ref{fig:fig_0_histogram} summarizes this quantitative comparison between BM25 and USE-QA on the development (dev) partitions of OpenBookQA and on a 10,000-query subset of ReQA SQuAD training partition. Here, we consider a model better than the other when it yields a better ranking for the correct justification. We draw two observations from this analysis:
{\flushleft {\bf (1)}} No approach is consistently better than the other. Overall, BM25 is at least as good as USE-QA in 293 queries out of 500 queries in OpenBookQA dev set, and  7603 queries out of 10000 randomly sampled queries in ReQA SQuAD train set. This is further motivation for a hybrid approach. 

{\flushleft {\bf (2)}} Importantly, there is immediate signal to differentiate between the two situations. When BM25 performs better than or similarly to USE-QA, the top BM25 score (after the softmax normalization) tends to be in the 0.8 to 1 range. In contrast, when there is little lexical overlap between question and justification indicated by low BM25 scores, e.g., below 0.2, USE-QA performs considerably better. This supports the intuition that USE-QA  can capture lexical differences between question and justification when present. 

\subsection{Hybrid Retrieval Model}

Motivated by the previous observations, we propose a hybrid evidence retrieval method that uses a routing classifier to direct an incoming question to either the BM25 retriever or the USE-QA retriever based on simple statistics that can be extracted efficiently. The key intuition behind our hybrid strategy is that we can estimate the optimal retrieval method
based on the top answers retrieved by traditional IR. In particular, if these answers receive a high score from the traditional IR method, it indicates that the current scenario is driven by lexical overlap, and traditional IR is likely to do better; the opposite is true otherwise.

We propose two variants for the routing classifier:

\label{sec:model_hybrid_retrieval}
{\flushleft {\bf Hybrid (1-param):}} This classifier uses a single parameter: a threshold on the normalized BM25 score of top document retrieved by the BM25 method.\footnote{We normalize this score by applying a softmax layer to the BM25 scores of the top $k$ ($k = 64$ in this paper) documents.}
If the top 1 score is higher than this threshold, the classifier routes the question to BM25; otherwise it sends it to USE-QA. This is a simple implementation of the intuition above -- if the top normalized BM25 score is high, then it is likely that there is a candidate document that has a large lexical overlap with the query, and which is probably a correct justification.
On the other hand, if the top normalized BM25 score is low, it is either because: (a) there is no document that has a large lexical overlap with the query, or (b) because there are multiple candidate documents that have high BM25 scores (and they are squished during normalization). In either of these scenarios BM25 is unlikely to identify the gold document, and, therefore, USE-QA should be selected.
The value of this threshold is determined by performing a grid search on the dev partition, with the threshold ranging from 0 to 1 with an interval of 0.1.


{\flushleft {\bf Hybrid (BM25):}} This classifier is a generalization of the above. That is, instead of relying solely on the top retrieved document, this classifier extracts features from the top $k$. In particular, 
for each query, we construct a feature vector $f$ and use a logistic regression (LR) classifier that takes $f$ to predict whether to use BM25 or USE-QA. The $i^{th}$ feature of $f$ is computed as $f_i=mean(S[0:2^i])$, where $S$ are the top BM25 scores ranked in the descending order (after softmax normalization). In this paper we use $i$ up to 6 (i.e., use up to top 64 BM25 scores). 
For example, feature 2 averages the BM25 scores of the top 4 documents retrieved by the traditional IR method. 
This strategy allows the classifier to take advantage of more documents when needed, but also focus on the top result(s) when they are sufficiently predictive. 

\section{Results}

In this section, we empirically evaluate the proposed evidence retrieval methods.
We use the mean reciprocal rank (MRR) score~\cite{voorhees1999proceedings} of the correct evidence sentence (or target document) in the test dataset as the evaluation measure.

Since ReQA SQuAD only provides training and developments partitions, we randomly sample 10,000 queries from the training data and use them for development, and use the original development set of ReQA SQuAD as test. ReQA NQ does not provide training/development/test partitions; for this dataset we use 5-fold cross-validation, sampling 10,000 queries from one fold as the development data in each split, and using the remaining folds as test. 

For all datasets, USE-QA is used without fine-tuning as proposed in \citet{ahmad2019reqa}. For the BERT retriever, we fine-tune it on the training data of OpenBookQA and ReQA SQuAD. For all hybrid retrievers, we tune their routing classifiers 
on the respective development partitions. 
For ReQA SQuAD, we further divide the development set into 5 splits (2,000 queries in each split) and tune 5 routing classifiers and evaluate them separately on the full test set to make sure the results are robust to different development sets. 

\subsection{Individual vs Hybrid Retrievers} 


Table \ref{tab:main_result} shows the MRR scores of the individual retrieval methods compared to the hybrid ones, on the three datasets.
We draw several observations from this table:



\begin{table}[ht]
\centering
\small
\begin{tabular}{lccccc}
\hline 
\textbf{ } & \textbf{OpenBook} & \textbf{ReQA}  & \textbf{ReQA}  \\ 
\textbf{ } & \textbf{QA} & \textbf{SQuAD}  & \textbf{NQ}  \\ \hline
BM25 & 0.522  & 0.645 & 0.293  \\
BERT & 0.557  & 0.260 & N/A  \\
USE-QA &  0.610  &  0.520  &   0.223 \\
BM25 + USE-QA & 0.550  & 0.647 & 0.290 \\
Hybrid (1-param)   & \textbf{0.611}  & 0.656 & 0.301  \\
Hybrid  (BM25)   & 0.596  & \textbf{0.657$^{*}$} & \textbf{0.303$^{*}$}  \\
Ceiling & 0.69 & 0.71 & 0.39 \\
\hline
\end{tabular}
\caption{Mean reciprocal rank (MRR) scores of the retrieval methods investigated on the three QA datasets. The {\em BM25 + USE-QA} method sums up the scores produced by BM25 and USE-QA, and uses that score for ranking. The {\em Hybrid (1-param)} method uses a routing classifier with a  single parameter: a threshold on the the BM25 score of the top document retrieved by BM25; the {\em Hybrid  (BM25)} uses features extracted from the top 64 documents retrieved by BM25. $^*$ indicates that  {\em Hybrid  (BM25)}  is statistically significantly better than both USE-QA and BM25 (bootstrap resampling with 10,000 iterations; $p$-value $< 10^{-5}$); on OpenBookQA there is no significant difference between {\em Hybrid  (BM25)} and USE-QA.
The {\em Ceiling} method always selects the best individual model (USE-QA or BM25) for each query by their ranking of the gold justification.
}
\label{tab:main_result}
\end{table}

\begin{table*}[ht]
\centering
\small
\begin{tabular}{lccccc}
\hline \textbf{ } & \textbf{OpenBookQA} & \textbf{ReQA SQuAD}  & \textbf{ReQA NQ}  \\ \hline
$n$ samples routed to BM25& 306  & 49,270 & 260,640  \\
$n$ samples routed to useQA & 194  & 7,860 & 59,845 \\ \hline
Samples w/ improved rankings vs. BM25 & 129  & 4,095 & 33,078  \\
Samples w/ worse rankings vs. BM25  & 47  & 3,257 & 25,562  \\
Samples w/ improved rankings vs. useQA  & 61  & 21,488 & 146,020 \\
Samples w/ worse rankings vs. useQA  & 93  & 8,640 & 84,839  \\

\hline
\end{tabular}
\caption{Routing statistics for the routing classifier that trains a logistic regression model using features extracted from the top 64 BM25 documents.}
\label{tab:routing_statistics} 
\end{table*}

{\flushleft {\bf (1)}} Most hybrid strategies outperform the individual retrieval methods, as well as the naive strategy that simply sums up the scores of two individual models, and uses the sum for ranking.
 {\em Hybrid (1-param)} and {\em Hybrid (BM25)} are statistically significantly better than BM25 and USE-QA on ReQA SQuAD and ReQA NQ under a bootstrap resampling significance analysis (10,000 samples, $p$-value $< 10^{-5}$). On OpenBookQA, {\em Hybrid (1-param)} and {\em Hybrid (BM25)} are statistically significantly better than BM25 under the same bootstrap resampling significance analysis, but there is no significant difference between the hybrid methods and USE-QA. 
 This demonstrates that transformer-based and IR-based methods capture complementary information, and the distinction of when to use one vs. another is learnable. Table~\ref{tab:routing_statistics} lists several runtime statistics of our best classifier, {\em Hybrid (BM25)}, which support this observation. The first two rows indicate that, indeed, the routing classifier uses both individual retrievers, with around 60\% (OpenBookQA) or 86\% (ReQA SQuAD) of questions being routed to BM25. The next four rows indicate that, on average, the hybrid approach improves over both individual methods especially on ReQA SQuAD and ReQA NQ. 

{\flushleft {\bf (2)}} While {\em Hybrid (BM25)} outperforms the simpler {\em Hybrid (1-param)}, the difference is not statistically significant.\footnote{10,000 samples, $p$-value $< 0.13$} This further suggests that simpler approaches work in this case. The routing decision can be approximated with a single parameter (a threshold on the BM25 score), applied to a single justification that is efficiently extracted by IR.\footnote{Supplemental material  \ref{supp:feature_analysis} has a more detailed discussion of what features to use.}

\begin{table}[ht]
\centering
\small
\begin{tabular}{l c c c}
\hline 
  & \textbf{OpenBook} & \textbf{ReQA} & \textbf{ReQA}  \\ 
  & \textbf{QA} & \textbf{SQuAD} & \textbf{NQ} \\ \hline
BM25 & 1.38  & 179.56 & 1547.56  \\
USE-QA &  19.74  &  3241.73  &  26722.63 \\
BM25 + USE-QA & 21.23  & 3476.99 & 28929.47 \\
Hybrid (1-param)  & 20.85  & 922.28& 9513.97  \\
Hybrid (BM25)   & 8.95  & 625.74 & 6565.21 \\
\hline
\end{tabular}
\caption{Runtime comparison of BM25, USE-QA and hybrid retrievers on the corresponding test partitions. All times are the total times in seconds on all test queries.}
\label{tab:time_comparison}
\end{table}

\subsection{Runtime Analysis}

A further advantage of our hybrid approach is improved runtime over neural methods, because a considerable number of queries are routed to a traditional, fast IR engine. To investigate this, 
we measure the processing time per query using BM25, USE-QA and various hybrid retrievers and calculate the total time usage of these retrieval methods. The processing time of each query is measured in the following way:
{\flushleft {\bf (1)}} For BM25, we measure the time of parsing the query, searching the top $k$ ($k = 1400$ for OpenBookQA, and 2000 for ReQA SQuAD and ReQA NQ) documents, and sorting the retrieved documents by the BM25 scores. 
{\flushleft {\bf (2)}} For USE-QA, we measure the time for query processing (including query tokenization and the embedding generation of USE-QA\footnote{The batch size is set to 1 when generating the embedding, for a fair comparison with BM25, and because in a real use case the queries may not arrive in batch.}), searching the top $k$ (1326 for OpenBookQA and 2000 for ReQA SQuAD) documents, and sorting them by the scores. We run this experiment using Tensorflow on Google Colab with GPU.
{\flushleft {\bf (3)}} For hybrid models, the processing time of each query is the sum of: (a) the BM25 processing time (2) the runtime of the routing classifier and (3) the processing time of USE-QA if USE-QA is selected for that query. 

In preliminary experiments, we found that the first few queries are processed slower due to the warming up of the various components. To control for this, we execute several warm-up queries before the actual time recording.

Table \ref{tab:time_comparison} shows the total processing time of all queries using different retrieval methods. The table indicates that USE-QA is more than 15 times slower than BM25 on all datasets.
Further, the hybrid approach reduces that gap while still allowing for the benefits of the neural IR when needed. For example, {\em Hybrid (BM25)} is 2.2 times faster than USE-QA in OpenBookQA, and 5.2 times faster in ReQA SQUAD.


\section{Conclusion}

This paper argues that, despite the exciting progress reported in the past couple of years, transformer network-based approaches such as BERT or USE-QA do not always outperform information retrieval methods such as tf-idf for evidence retrieval for QA. We validate this observation both with a direct empirically analysis, and, more importantly, with a lexical probing task where two probes were trained to predict words in the gold evidence texts. We observed that the first probe, 
 which was trained on the tf-idf vector of the query, tends to predict words that exist in the original query (thus emphasizing lexical overlap), whereas the second probe, trained on top of the query's neural embedding, predicts more words in the evidence text that do not exist in the query (thus bridging the lexical differences between these texts).

Learning from this analysis, we introduced a routing classifier that learns when to direct incoming questions to traditional or neural IR methods for evidence retrieval. The routing classifier is trained using very simple statistics, which can be extracted from the top candidate evidence sentences produced by a traditional IR model. We demonstrated that this hybrid evidence retrieval generally performs better than either individual retrieval strategy on three QA datasets. Further, we showed that this routing classifier can be approximated with nearly the same performance with a 1-parameter model (a threshold over the IR score of the top evidence sentence retrieved by BM25), which can drastically simplify real-world applications of our approach. 
Lastly, we show that our routing classifier is considerably faster than USE-QA, with runtime improvements of up to five times. 


\newpage
\bibliography{emnlp2020}
\bibliographystyle{acl_natbib}

\clearpage
\newpage
\appendix

\section{Supplemental Material}
\label{sec:supplemental}

\subsection{Probe Experiment Setup}
\label{sec:supp_probe}
{\flushleft {\bf Data Generation:}} For both probe tasks, we lemmatize all queries and documents using NLTK to build the vocabulary. Stop words are then removed from the vocabulary (so that the evaluation only considers the predictions of the probe model on those non-stop words, and the probe will not be trained to generate stop words). On ReQA SQuAD, we limit the vocabulary size by removing all tokens that appear less than 4 times in all paragraphs. 

The prediction task operates over terms, i.e., unique words, rather than the original tokens in the queries and facts.
For example if the tokens in the query are ``good'', ``human'', ``is'', and ``good'', the query terms considered will be ``good'' and ``human'' (one of the ``good'' tokens is removed due to duplication, and ``is'' is removed because it is a stop word).

{\flushleft {\bf Loss Function:}} The output size of the linear probe is $N_v$, which is the vocabulary size. Each output element of the linear probe layer is the predicted probability of that term showing up in fact/query. For example, the vocabulary is ["egg", "apple", "banana"], and the output of the linear probe is [0.1, 0.3, 0.9]. This means the probability of "egg" predicted by the query embedding is 0.1, and the probability of "banana" is 0.9. Note that the sum of all probabilities is not necessarily one (i.e., the output probabilities are independent of each other), because the query embedding might encode multiple terms. Every number of the output of the linear probe gets a binary cross entropy loss. We do not use softmax over all predicted values because that will enforce the classifier to predict only one term. Instead, we want the linear probe to predict all terms. Therefore we choose sigmoid function and binary cross entropy loss. For each query, the gold terms to be predicted are accompanied by the same number of randomly sampled negative terms (terms that should not be predicted). For example, if the terms to be predicted are "good" "cat", the negative terms might be "yellow" "noodle", and the labels for the terms ["good" ,"cat", "yellow", "noodle"] would be [1, 1, 0, 0]. Then the linear probe is trained using this label. Other nodes of the linear probe will not get supervision signals. This is to make sure the updates of the weights are not extremely imbalanced during training. 

{\flushleft {\bf Training:}} We use an Adam optimizer with the learning rate of 0.001. The final performance is achieved by the model in the epoch that has the lowest total evaluation loss. 

\subsection{Fine-tuning BERT for Retrieval}
\label{sec:supp_finetune_bert}

{\flushleft {\bf Encoding:}} We use two separate BERT ($bert_d$ for document and $bert_q$ for query) models to encode the query and the documents separately. Due to the limitation of computational resources, we limit the max length of the input documents to be 256 tokens. We set all segment masks to 1 in encoding. 
{\flushleft {\bf Loss Function:}} We use negative sampling to train the retriever. In each batch of document embeddings, one is the gold document embedding and the others are the non-relevant document embeddings. We multiple the query embedding with all document embeddings in the batch, yielding one relevance scores for each document in the batch $R = [r_1, r_2, ..., r_N]$. The final loss is the cross entropy loss on top of the softmax-normalized relevance scores $CrossEntropy(softmax(R), target)$. 

On OpenBookQA, the batch size is 11 (1 gold document and 10 non-relevant documents). On ReQA SQuAD, the batch size is 5 (1 gold document and 4 non-relevant documents) due to the memory limitation of our device.
{\flushleft {\bf Training:}} We use the Adam optimizer and set the learning rate to be $1e-5$. On OpenBookQA we fine-tune it for 4 epochs and on ReQA SQuAD we tune it for 2 epochs. 
{\flushleft {\bf Evaluation:}} We train the BERT retriever using 5 random seeds. Under each random seed, we randomly sample 2,000 questions from the training set to serve as the development set and use the original dev set as the test set. The final results in the table are the test MRR achieved by the best-performing model on dev under each seed. 

\newpage
\subsection{Feature Analysis of the Hybrid (BM25) Retriever}
\label{supp:feature_analysis}

To investigate whether using other features help the hybrid classifier, we modified the routing classifier to consider the top $k$ justification scores of BM25 and USE-QA.
In particular, {\em top{$a_1$, $a_2$, ..., $a_k$} Hybrid ($M$)} means the LR classifier uses up to top k scores of Method $M$ as described in Section \ref{sec:model_hybrid_retrieval}. In contrast,  {\em top $k$ Hybrid ($M$)} means the LR classifier uses merely the average top k scores of Method $M$. Table~\ref{tab:routing_topk_analysis} summarizes these results.  

The table indicates that {\em Hybrid (BM25)} is improved very marginally on ReQA SQuAD and ReQA NQ by including top USE-QA scores (row 1 and row 6 in Table \ref{tab:routing_topk_analysis}).

\begin{table*}[ht]
\centering
\small
\begin{tabular}{lccccc}
\hline \textbf{ } & \textbf{OpenbookQA} & \textbf{SQuAD}  & \textbf{NQ} \ \\ \hline
top \{1, 2, 4, 8, 16, 32, 64\} Hybrid(BM25, USE-QA)   & 0.596  & 0.658 & 0.305    \\
top 1 Hybrid (BM25, USE-QA)   & 0.597  & 0.658 & 0.302  \\
top 4 Hybrid (BM25, USE-QA)    & 0.551  & 0.648 & 0.303  \\
top 16 Hybrid (BM25, USE-QA)     & 0.522  & 0.645 & 0.293 \\
top 64 Hybrid (BM25, USE-QA)      & 0.522  & 0.645 & 0.293 \\ \hline
top \{1, 2, 4, 8, 16, 32, 64\} Hybrid (BM25)   & 0.596  & 0.657 & 0.303   \\
top 1 Hybrid (BM25)     &  0.598 &  0.656 & 0.299  \\
top 4 Hybrid (BM25)       & 0.548  & 0.648 & 0.302  \\
top 16 Hybrid (BM25)       & 0.522  & 0.645 & 0.293 \\
top 64 Hybrid (BM25)       & 0.522  & 0.645 & 0.293 \\ \hline
top \{1, 2, 4, 8, 16, 32, 64\} Hybrid (USE-QA)  & 0.522  & 0.645 & 0.293  \\
top 1 Hybrid (USE-QA)       & 0.522  & 0.645 & 0.293  \\
top 4 Hybrid (USE-QA)       & 0.522  & 0.645 & 0.293  \\
top 16 Hybrid (USE-QA)       & 0.522  & 0.645 & 0.293 \\
top 64 Hybrid (USE-QA)       & 0.522  & 0.645 & 0.293 \\ 
\hline
\end{tabular}
\caption{Use top k features for the linear regression routing classifier}
\label{tab:routing_topk_analysis} 
\end{table*}

\begin{figure*}
\centering
\includegraphics[width = 0.7\textwidth]{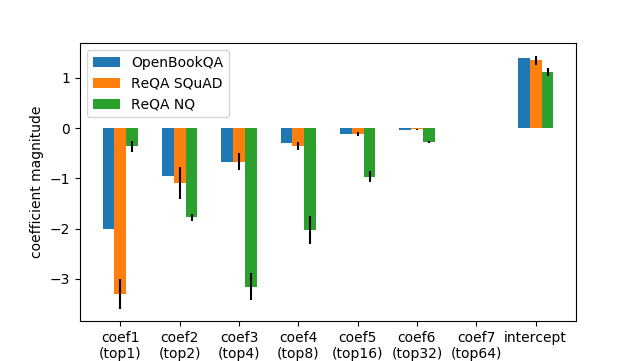}
\caption{\footnotesize Histograms of the coefficents of the logistic regression classifier in Hybrid (BM25) model. The x-axis also indicates which feature the coefficient corresponds with. For example, coefficient 2 corresponds with the average of top 2 BM25 scores. The standard deviations of the coefficients on ReQA SQuAD and ReQA NQ are represented by the small black bars. }
\label{fig:fig_4_feature_analysis}
\end{figure*} 
Figure \ref{fig:fig_4_feature_analysis} shows the magnitude of the coefficients of the Logistic Regression (LR) classifier in a Hybrid (BM25). We use label 1 to indicate ``choosing USE-QA'' and 0 to indicate ``choosing BM25'' for the LR classifier. For OpenBookQA and ReQA SQuAD, a large top BM25 score is strongly related to choosing BM25, whereas on ReQA NQ, a large "average of top 4 BM25 scores" is strongly related to choosing BM25.

\newpage
\subsection{Example Outputs}
\label{supp_exp_output}

\begin{table*}[t]
    \centering
    \small
    \begin{tabular}{ p{0.3cm} p{5cm} p{4.7cm} p{0.7cm} p{0.7cm}p{0.7cm}p{0.7cm}} \hline
      $\#$ &query & gold fact & BM25 MRR & USE-QA MRR & hybrid MRR & type\\ \hline
      1 &  What would happen when balloons heat up? they get bigger & as heat increases , a flexible container containing gas will expand & 0.02 & 0.33 & 0.33 & TP\\ 
      2 & There is most likely going to be fog around: a marsh  & fog is formed by water vapor condensing in the air & 1.0 & 0.33 & 0.33 & FP \\
      3 & The summer solstice in the northern hemisphere is four months before October & the summer solstice is on June 21st in the northern hemisphere & 1.0 &0.25 &1.0 & TN \\
      4 & An example of conservation is avoiding the use of gasoline & An example of conservation is not using fossil fuel & 0.5 & 1.0 & 0.5 &FN\\ \hline
      
    \end{tabular}
    \caption{Examples of the classifier's choices. True Positives (TP): the routing classifier decides to use USE-QA, and USE-QA is indeed better; False Positives (FP): the routing classifier decides to use USE-QA, but BM25 is actually better; True Negatives (TN): the routing classifier decides to use BM25, and BM25 is indeed better; False Negatives (FN): the routing classifier decides to use BM25, but USE-QA is actually better.}
    \label{tab:classifier_examples}
\end{table*}

Table \ref{tab:classifier_examples} shows some examples of the behavior of the routing classifier on OpenBookQA. In example 1, the gold justification has very little word overlap with the query. Therefore, the ranking of BM25 is very low. However, the tokens ``container'' and ``bigger'' in the justification are semantically related to the tokens ``balloon'' and ``increase'', so that the gold justification is ranked high by USE-QA. In addition, the hybrid classifier is able to correctly route this question to USE-QA. 

In example 3, the gold justification has considerable word overlap with the query, therefore its BM25 ranking is very high. The routing classifier is also able to choose to use BM25.

In examples 2 and 4 the routing classifier does not function correctly. In example 2, we conjecture that the small word overlap ratio confuses the routing classifier to direct the question to USE-QA.
In example 4, multiple facts (not listed) have high BM25 scores, which incorrectly influence the routing decision.

\end{document}